\documentclass{elsart3}

\usepackage{amssymb}
\usepackage{epsfig}
\usepackage{amsmath}

\begin{document}

\begin{frontmatter}

% Title, authors and addresses

% use the thanksref command within \title, \author or \address for footnotes;
% use the corauthref command within \author for corresponding author footnotes;
% use the ead command for the email address,
% and the form \ead[url] for the home page:
% \title{Title\thanksref{label1}}
% \thanks[label1]{}
% \author{Name\corauthref{cor1}\thanksref{label2}}
% \ead{email address}
% \ead[url]{home page}
% \thanks[label2]{}
% \corauth[cor1]{}
% \address{Address\thanksref{label3}}
% \thanks[label3]{}

  \title{Tight-binding model of spin-polarized tunnelling in
    (Ga,Mn)As-based structures}

% use optional labels to link authors explicitly to addresses:
% \author[label1,label2]{}
% \address[label1]{}
% \address[label2]{}
  \author[addr1]{P. Sankowski}
  \author[addr1]{P. Kacman}
  \author[addr2]{J. Majewski}
  \author[addr1,addr2,addr3]{T. Dietl}
  \address[addr1]{Institute of Physics, Polish Academy of Sciences,
    al. Lotnikow 32/46, 02 668 Warszawa, Poland}
  \address[addr2]{Institute of Theoretical Physics, Warsaw University, ul. Hoza 69, 00 681 Warszawa, Poland}
  \address[addr3]{ERATO Semiconductor Spintronics Project of JST, al. Lotnikow 32/46, 02 668 Warszawa, Poland}
  
\begin{abstract}
  The Landauer-B\"uttiker formalism combined with the tight-binding
  transfer matrix method is used to describe the results of recent
  experiments: the high tunneling magnetoresistance (TMR) in
  (Ga,Mn)As-based trilayers and highly polarized spin injection in
  p-(Ga,Mn)As/n-GaAs Zener diode.  For both TMR and Zener spin current
  polarization, the calculated values agree well with those observed
  experimentally. The role played in the spin dependent tunneling by
  carrier concentration and magnetic ion content is also studied.
\end{abstract}

\end{frontmatter}

% main text
\section{Introduction}
Efficient spin injection is a fundamental prerequisite for
construction of spintronic devices. On the other hand, the tunneling
magnetoresistance (TMR) effect, examined in a pioneering work by
Julli\'ere \cite{Julliere}, has found already many applications in,
e.g., magnetic field sensors and magnetic random access memories,
where the polycrystalline transition metals are usually employed as
ferromagnetic layers.  Both effects have been observed in
(Ga,Mn)As-based structures.  The first observation of a high (of about
75\%) TMR effect was reported for a trilayer structure
(Ga,Mn)As/AlAs/(Ga,Mn)As \cite{tanaka}.  Recently, TMR was observed in
(Ga,Mn)As/GaAs/(Ga,Mn)As structures, \cite{Matt03,chiba} reaching
about 300\% at low temperatures in devices, in which (Ga,Mn)As
contained about 8\% of Mn \cite{chiba}. Furthermore, peculiar
behavior of TMR was observed in nanoconstrictions \cite{Gidd05}
as well as when the holes in (Ga,Mn)As were at the localization boundary
\cite{Rust05}. Also, it has been demonstrated that highly spin
polarized electron current, with polarization reaching 80\%, can be
obtained from a p-(Ga,Mn)As/n-GaAs Zener diode \cite{vandorpe}.

\section{Theoretical Model}

To describe the spin dependent processes
we employ the  Landauer-B\"uttiker formalism for coherent tunneling.
The  transmission coefficients, needed in this approach,
are determined in terms of the extended transfer-matrix method
within the tight-binding framework that takes into account spin
dependent terms. The tight-binding Hamiltonian matrix is composed of
three parts: the left and right leads, and the middle
region, where the incoming Bloch waves from the left lead are
scattered into outgoing Bloch states of the right lead. By solving the
Schr\"{o}dinger equation for the tight-binding Hamiltonian, we determine the transfer
coefficients $t_{k_{\bot,i} \to k_{\bot,j}}(E, \mathbf{k}_{\|})$,
which describe the probability of tunneling from the incoming state
$k_{\bot,i}$ to the outgoing state $k_{\bot,j}$ for given electron
energy $E$ and wave vector parallel to the surface
$\mathbf{k}_{\|}$~\cite{Christian}. The tunneling current
$\mathbf{j}$ is given by,
\begin{align}
  \label{equation-lb}
  \mathbf{j} =& \frac{-e}{4\pi^3 \hbar} \int_{BZ} d^2k_{\|}dE
  \sum_{\substack{ k_{\bot,i},k_{\bot,j} \\ v_{\bot,i},v_{\bot,j}>0}}
%  \!\!\!\!
  \\ \nonumber
  &
  \left[f_L(E) - f_R(E)\right]
\left|
    t_{k_{\bot,i} \to k_{\bot,j}}(E, \mathbf{k}_{\|})
  \right|^2 \frac{v_{\bot,j}}{v_{\bot,i}},
\end{align}
where $f_L$ and $f_R$ are the electron Fermi distributions in the left
and right interface and $v_{\bot,i}$ are the group velocities of the
corresponding Bloch states.

In order to construct the empirical tight-binding Hamiltonian matrix
for the heterostructure we have to start from the description of the
constituent materials. To describe the band structure of bulk GaAs
we adopt the $sp^3d^5s^*$ tight-binding parametrization, with the
spin-orbit coupling included, as proposed by Jancu {\em et
al.} \cite{jancu}. This model reproduces correctly the effective
masses and the band structure of GaAs in the whole Brillouin zone, in
agreement with the results obtained by empirically corrected
pseudopotential method. The parametrization includes only the nearest
neighbor (NN) interactions. For each anion and cation 20 orbitals are
used - hence, with each GaAs layer (0.28~nm) of the structure the size
of the tight-binding matrix increases by 40. The same set of orbitals
is used to describe the (Ga,Mn)As layer.  It should be pointed out
that the $d$ orbitals used in our $sp^3d^5s^*$ parametrization are not
related to the Mn ions incorporated into GaAs. The presence of Mn ions
in (Ga,Mn)As is taken into account only by including the $sp$-$d$
exchange interactions within the virtual crystal and mean-field
approximations.  The values of the exchange constants are determined
by the experimentally obtained spin splittings: $N_0\alpha=0.2$~eV of
the conduction band and $N_0\beta=-1.2$~eV of the valence
band \cite{okabayashi}. The other parameters of the model for the
(Ga,Mn)As material are taken to be the same as for GaAs -- this is
well motivated because the valence-band structure of (Ga,Mn)As with
small fraction of Mn was shown to be quite similar to that of GaAs
\cite{okabayashi}. We construct the tight binding matrix for the
heterostructure taking for each double layer of the
structure the description of the corresponding bulk material. The NN
interactions between GaAs and (Ga,Mn)As are described by the same
parameters as the interactions in bulk GaAs. Consequently, the valence
band offset between (Ga,Mn)As and GaAs originates only form the spin
splitting of the bands in (Ga,Mn)As.

The Fermi energy in the constituent materials is determined by summing up
the occupied states over the entire Brillouin zone. The number of
occupied states is determined by the assumed carrier concentration in
the material. Our calculations of the dependence of Fermi energy on
hole concentration are consistent with the corresponding results
presented previously \cite{dietl-01}. It should be noted that the Fermi
energy in Ga$_{1-x}$Mn$_{x}$As depends crucially on the hole concentration,
whereas very little on the Mn content $x$.

\section{Results}

\subsection{Tunneling Magnetoresistance}

\begin{figure}
  \epsfig{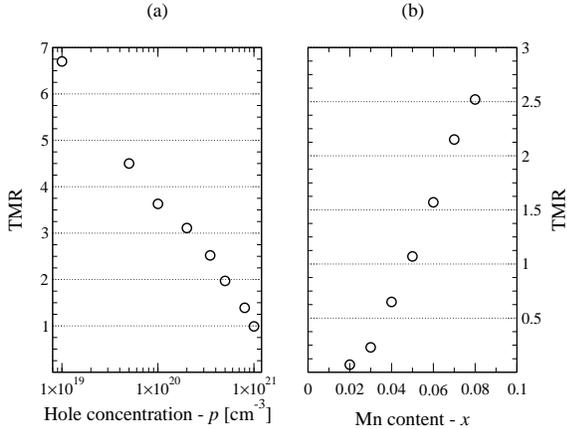}
  \caption{TMR in p-Ga$_{1-x}$Mn$_{x}$As/n-GaAs as a function of: (a) hole
  concentration $p$ (for $x=0.08$); (b) Mn content (for $p= 3.5 \times
  10^{20}$~cm$^{-3}$). The bias applied to the structure is $V =
  0.05$~V.}
 \label{figure-1}
\end{figure}

In our calculations of the TMR effect we consider the structure
containing three layers. The two half-infinite leads are
build of the ferromagnetic p-type Ga$_{1-x}$Mn$_{x}$As. The middle
scattering region is composed of the non-magnetic GaAs, which forms a
barrier for the holes. We compare the tunneling currents in two configurations, i.e., with
parallel (ferromagnetic -- FM) and the antiparallel (antiferromagnetic --
AFM) alignments of the magnetizations in the leads. We define the
tunneling magnetoresistance as
\[
TMR = \frac{I_{FM} - I_{AFM}}{I_{AFM}},
\]
where $I_{FM}$ and $I_{AFM}$ are the currents in the FM and AFM
configurations, respectively. In Fig.~\ref{figure-1} (a)
the obtained TMR values, for a given  (8\%) Mn content and a set of
different hole concentrations in the FM layers,
are plotted as a function of the applied bias.
As shown, TMR depends strongly on the hole
concentration. As TMR is primarily determined by spin
polarization of the carriers at the Fermi level, the higher hole
concentration the smaller is spin polarization at the
Fermi level at given Mn
spin polarization. For $p = 3.5 \times 10^{20}$~cm$^{-3}$,
which is the typical hole concentration in (Ga,Mn)As
samples with a high Mn content \cite{Yu-apl02}, the TMR of about 250\% is
obtained.

Because of self-compensation,  the hole concentration depends rather weakly on $x$
-- thus, we have calculated the TMR for
different $x$ in the magnetic layers, while keeping the hole concentration 
constant, $p = 3.5 \times 10^{20}$~cm$^{-3}$.
The results of such computations are presented in Fig.~\ref{figure-1}(b).

% \begin{figure}
%   \epsfig{file=gaga_0.250_all_c.eps,scale=0.3}
%   \caption{TMR in Ga$_{1-x}$Mn$_{x}$As/GaAs/Ga$_{1-x}$Mn$_{x}$As
%     vs. the applied bias for structures with different content of Mn ions in
% the magnetic layers. The hole concentration in the
%     magnetic regions is assumed $p= 3.5 \times 10^{20}$~cm$^{-3}$ for all values.}
%  \label{figure-2}
% \end{figure}

As seen, our simple model reproduces fully the experimental
data: for structures with 8\% of Mn we obtain the TMR of the order of
250\%, as observed recently by Chiba {\it et al.}~\cite{chiba}; for
4\% of Mn the calculations lead to the TMR of the order of 60\%, in
perfect agreement with the observations of Tanaka and
Higo~\cite{tanaka} and Mattana {\it et al.}~\cite{Matt03}.  Therefore,
our calculations seem to suggest that for obtaining a high TMR, large
exchange splittings, i.e., high content of magnetic ions is needed.
Unfortunately, the presented in Fig.~\ref{figure-1} dependence
suggests that the attempts to increase the hole concentration in
(Ga,Mn)As, in order to obtain higher Curie temperature, may result in
a reduced TMR value.

% \begin{figure}
%   \epsfig{file=gaga_0.08_width.eps,scale=0.3}
%   \caption{.}
%  \label{figure-7}
% \end{figure}

\subsection{Zener Diode}
As a second application of the developed formalism for spin-dependent
tunneling, we consider the Zener diode in which high polarization of
the spin current has been recently observed \cite{vandorpe}. Alas, in
approaches involving transfer matrix formalism, computational
constraints hinder the simulations of the spin-dependent tunneling
through the whole device used in the experiments in Ref.
\cite{vandorpe}.  Therefore, we consider the simpler
p-Ga$_{1-x}$Mn$_x$As/n-GaAs structure with relatively narrow depletion
region consisting of 4 double-layers. Albeit simplified, such
structure captures the essential physics concerning tunneling of
electrons from the spin-polarized valence band of (Ga,Mn)As to the
GaAs conduction band.  Moreover, this approach can provide
quantitative information on spin polarization of the current, even
though it overestimates necessarily the tunneling current.
Simulations which take into account a more realistic description of
the depletion region are presented elsewhere \cite{pol}.

In our computation, we
assume that the magnetization vector is in $110$ direction and we
evaluate the spin current polarization in respect to this direction.
We use Eq.~\eqref{equation-lb} to compute separately the currents
$\mathbf{j}_{\uparrow}$ of spin up and $\mathbf{j}^{\downarrow}$ of
spin down electrons. The spin current polarization $P_j$ is defined as
follows:
\[
P_j = \frac{\mathbf{j}^{\uparrow} - \mathbf{j}^{\downarrow}}
{\mathbf{j}^{\uparrow} + \mathbf{j}^{\downarrow}}.
\]

\begin{figure}
  \epsfig{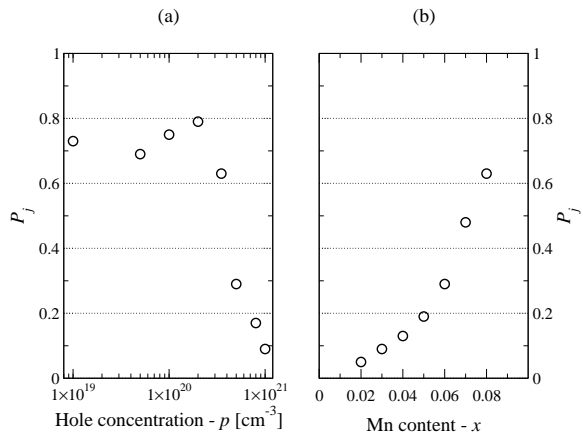}
  \caption{Spin current polarization $P_j$ in p-Ga$_{1-x}$Mn$_{x}$As/n-GaAs
as a function of: (a) hole concentration $p$ (for $x=0.08$); (b) Mn content (for
$p= 3.5 \times 10^{20}$~cm$^{-3}$). The bias applied to the structure is $V = 0.05$~V.}
 \label{figure-3}
\end{figure}

We assume that the electron concentration
is $n = 10^{19}$~cm$^{-3}$ as indicated by the experimental results in
Ref.~\cite{vandorpe}. The dependencies of $P_j$ on the hole
concentration $p$  and Mn content $x$ are depicted in Fig.~\ref{figure-3}.

Similarly to TMR, the presented in Fig.~\ref{figure-3} results show a
strong decrease of the tunneling current spin polarization with the
increase of hole concentration.  The spin injection in the Zener
diode, again like in TMR, depends crucially on the content of magnetic
ions in the Ga$_{1-x}$Mn$_{x}$ layer. For $x=0.08$ we obtain the spin
current polarization to be of the order of 60\%, what agrees very well
with the experimental observations \cite{vandorpe}.

The calculated TMR and spin current polarization in the Zener diode
both decrease rapidly with the applied bias voltage, as observed in
the experiments. This "zero-bias anomaly" has been observed before in
many planar metal-insulator-metal tunnel junctions, but is still far
from being completely understood. The results of our tight-binding
model suggest that it is primary due to the band structure effects.

\section{Summary}
We have analyzed the spin-polarized tunneling in (Ga,Mn)As-based
structures employing a tight-binding model together with the
Landauer-B\"{u}ttiker formalism.  Our studies reproduce quantitatively
the recently observed high TMR in (Ga,Mn)As/(Al,Ga)As/(Ga,Mn)As
trilayers and large spin polarization of the injected current in a
(Ga,Mn)As/(Al,Ga)As spin-LED. The model describes as well the strong
dependence of spin injection on the applied bias voltage. It should be
pointed out that our calculations are not self-consistent at the
moment.  However, in contrast to the standard $k\cdot p$ method
\cite{petukhow,brey}, the scattering formalism based on the
tight-binding scheme takes into account all the effects resulting from
the electric field in the depletion zone, in particular, the Rashba
and Dresselhaus terms that are essential for tunneling. These features
make our approach particularly suited for studying phenomena related
to spin-polarized tunneling.

\section{Acknowledgments}
We thank F. Matsukura and H. Ohno for valuable discussions. 
This work was supported by the Polish
Ministry of Science, Grant PBZ-KBN-044/P03/2001. The calculations
were carried out at ICM in Warsaw.

% The Appendices part is started with the command \appendix;
% appendix sections are then done as normal sections
% \appendix

% \section{}
% \label{}

\end{document}